\documentclass[a4paper,11pt]{article}
\pdfoutput=1 
\usepackage{jheppub} 
\usepackage[T1]{fontenc} 
\usepackage{amssymb}
\usepackage{amsthm}
\usepackage{amsmath}
\usepackage[usenames,dvipsnames]{xcolor}
\usepackage{epsfig}
\usepackage{dcolumn}
\usepackage{tikz}
\usetikzlibrary{shapes.geometric, arrows}
\usepackage{upgreek}
\usepackage{setspace}
\usepackage{enumitem}
\usepackage{array,multirow,bigdelim,arydshln}
\usepackage{appendix}
\usepackage{xparse}
\usepackage{graphicx} 
\usepackage{mathtools}
\usepackage{physics} 
\usepackage[section]{placeins}
\usepackage[utf8]{inputenc}
\hypersetup{
	colorlinks,
	urlcolor=Maroon,
	linkcolor=Maroon,
	citecolor=Maroon
}

\usepackage{float}

\NewDocumentCommand{\binomial}{omm}
{%
	\genfrac(){0pt}{}{#2}{#3}%
	\IfValueT{#1}{_{\!#1}}%
}
\NewDocumentCommand{\eulerian}{omm}
{%
	\genfrac<>{0pt}{}{#2}{#3}%
	\IfValueT{#1}{_{\!#1}}%
}

\usepackage{underscore}

\usepackage{latexsym}
\usepackage{tikz}

\theoremstyle{plain}

\newtheorem{thm}{Theorem}
\newtheorem{defn}[thm]{Definition}
\newtheorem{cor}[thm]{Corollary}
\newtheorem{prop}[thm]{Proposition}

\newtheorem{rem}[thm]{Remark}
\newtheorem{example}[thm]{Example}

\newtheorem{conjecture}[thm]{Conjecture}

\usepackage{latexsym}

\title{Planar kinematic invariants, matroid subdivisions and generalized Feynman diagrams}

 \author{Nick Early}

\affiliation{Perimeter Institute for Theoretical Physics, Waterloo, Ontario}

\emailAdd{earlnick@gmail.com}

\abstract{In recent work of Cachazo, Guevara, Mizera and the author, a generalization of the biadjoint scattering amplitude $m^{(k)}(\mathbb{I}_n,\mathbb{I}_n)$ was introduced as an integral over the moduli space of $n$ points in $\mathbb{CP}^{k-1}$, with value a sum of certain rational functions on the kinematic space $\mathcal{K}_{k,n}$.  It was shown there for $m^{(3)}(\mathbb{I}_6,\mathbb{I}_6)$ and later by Cachazo and Rojas that collections of poles appearing in $m^{(3)}(\mathbb{I}_7,\mathbb{I}_7)$ are compatible exactly when they are dual to collections of rays which generate the maximal faces of a polyhedral complex known as the (nonnegative) tropical Grassmannian.  

In this note, we derive a remarkable planar basis for the space of generalized kinematic invariants which coincides in the case $k=2$ with usual standard planar multi-particle basis for the kinematic space.  We implement in Mathematica the action on formal linear combinations of planar matroid subdivisions of a boundary operator which, together with the planar basis, determines compatibility for any given poles appearing in the expansion of $m^{(k)}(\mathbb{I}_n,\mathbb{I}_n)$, by computing a certain combinatorial non-crossing condition on the second hypersimplicial faces $\Delta_{2,n-(k-2)}$ of $\Delta_{k,n}$.  The algorithms are implemented in an accompanying Mathematica notebook and are evaluated on existing tables of rays, in the form of tropical Plucker vectors, to tabulate the finest planar subdivisions of $\Delta_{3,8},\Delta_{3,9}$ and $\Delta_{4,8}$, or equivalently the set of maximal cones for the corresponding nonnegative tropical Grassmannians.

}

\begin{document} 
\maketitle
\flushbottom

\section{Introduction}
This note aims to supplement our previous work \cite{Early19WeakSeparationMatroidSubdivision} and to make available techniques which may be useful in studying combinatorially the expansion of the generalized biadjoint amplitudes $m^{(k)}(\mathbb{I}_n,\mathbb{I}_n)$ introduced in \cite{TropGrassmannianScattering} and studied subsequently in \cite{CachazoRojas,TropicalGrassmannianCluster Drummond,CachazoBorges,SoftTheoremtropGrassmannian,CachazoSingularSolutions,Henke}.  In our approach there are two main objects.  The first layer consists of a family of piecewise continuous\footnote{Unless otherwise stated, all functions $h$ on the hypersimplex shall be assumed piecewise continuous.} functions on a hypersimplex\footnote{Recall that hypersimplices are the convex polytopes $\Delta_{k,n} = \left\{x\in\lbrack 0,1\rbrack^n: \sum_{j=1}^n x_j=k\right\}$,see Equation \eqref{eq:deltaJ}.} $\Delta_{k,n}$ whose curvatures are identically zero over a certain alcove triangulation, i.e. one that is compatible with a given cyclic order, as in \cite{AlcovedPolytopes}.  We derive a subfamily of these which, when evaluated on the vertices of $\Delta_{k,n}$, have the following important properties: (1) they are in bijection with ``nonfrozen\footnote{Labeled by subsets not consisting of a single cyclic interval}'' vertices, (2) they are all poles of $m^{(k)}(\mathbb{I}_n,\mathbb{I}_n)$, (2) modulo zero curvature functions they dualize to give a planar basis for linear functionals on the kinematic space, and (3) expanding any pole in the planar basis makes possible a combinatorial criterion for the compatibility of any two poles: (4) given a pair of functions which have zero curvature over the maximal cells of some matroid polytopes, compute these curvatures, restrict them to the second hypersimplicial faces $(\simeq \Delta_{2,n-(k-2)})$ of $\Delta_{k,n}$, and check whether the usual four-term Steinmann relations hold on pairs of 2-block set partitions, see \cite{Steinmann1} and in particular the review article \cite{Streater75}.

In a Mathematica notebook which accompanies the arXiv submission we implement (4) to enumerate the maximal cones in $\text{Trop}_+(k,n)$ for $(k,n)\in\{(3,6),(3,7),(3,8),(3,9),(4,8)\}$, obtaining the same enumeration which was found in \cite{CachazoPlanarCollections}.  See Appendix \ref{sec: Feynman Diagram Enumeration} for a summary of the results.

By analogy with the case $k=2$ the expansion of $m^{(k)}(\mathbb{I}_n,\mathbb{I}_n)$ for $k\ge3$ were called generalized Feynman diagrams, and their study, as objects of independent interest, was initiated in \cite{CachazoBorges} for $k=3$ using collections of trees, and then to $k\ge4$ in \cite{CachazoPlanarCollections} using arrays of Feynman diagrams.  

In Section \ref{section:height functions} we derive a new planar basis of linear functionals on $\mathbb{R}^{\binom{n}{k}}$ denoted $\eta_J$, for certain nonfrozen $k$-element subsets $J$ of $\{1,\ldots, n\}$.  These give rise to a planar basis of the kinematic space $\mathcal{K}_{k,n}$ when it is viewed as a subspace of $\mathbb{R}^{\binom{n}{k}}$.  The basis of functionals $\eta_J$ on the kinematic space are labeled by sets $J$ whose indices do not form a cyclic interval in $\{1,\ldots, n\}$.  

Our perspective is that the biadjoint amplitude, $m_n^{k}(I,I),$ admits a series of inner approximations defined by
\begin{eqnarray}\label{eq: restricted amplitude0}
m^{(k)}_{n,\ \text{inner}} = \sum_{\mathcal{C} = \{ J_1,\ldots,J_{N_k}\}}\frac{1}{\eta_{J_1}\cdots \eta_{J_{N_k}}},
\end{eqnarray}
in correspondence with certain polyhedral subcomplexes of the nonnegative tropical Grassmannian $\text{Trop}_+ G(k,n)$.  Here $N_k = (k-1)(n-k-1)$, and the sum is over all compatible collections $\mathcal{C} = \{J_1,\ldots, J_{(k-1)(n-k-1)}\}$ of planar basis elements $\eta_{J}$, where the $J_i$ are nonfrozen $k$-element subsets of $\{1,\ldots, n\}$.  We construct the first inner approximation completely using the planar basis, and then skip all the way to compute the whole amplitude using ray data which was kindly shared with us.  Compatible collections consist of (non cyclically-consecutive) $k$-element subsets of $\{1,\ldots, n\}$ which satisfy a pairwise non-crossing condition known as \textit{weak separation}: if $e_J := \sum_{j\in J} e_J$ with $\vert J \vert = k$, then for any pair $J_a,J_b$ the difference $$e_{J_1}-e_{J_2}$$
must avoid the pattern 
$$e_a-e_b+e_c-e_d$$
given the cyclic orientation $a<b<c<d$.  In the case $k=2$, then weak separation becomes a special case of the Steinmann relations\footnote{This is in the sense presented in the review article \cite{Streater75}; see \cite{Steinmann1,Steinmann2} for the original texts in German.} on transverse (affine) hyperplanes.  See also \cite{LiuNorledgeOcneanu,NorledgeOcneanu} for connections to Hopf algebras.

By summing over all such collections we obtain the first \textit{inner approximation} to $m^{(k)}(\mathbb{I}_n,\mathbb{I}_n)$, given in Equation \eqref{eq: restricted amplitude0}.

The justification for the term ``inner approximation'' becomes more clear when stated geometrically: it means, by Theorem \ref{thm: weakly separated matroid subdivision blade}, that we are summing over the all refinements of a particular family of matroid subdivisions known as \textit{multi-splits}, where all maximal cells are required to be positroid polytopes.  Equivalently, the sum is over maximal faces of the correspondingly labeled polyhedral subcomplex of the nonnegative tropical Grassmannian.

\begin{figure}[h!]
	\centering
	\includegraphics[width=0.85\linewidth]{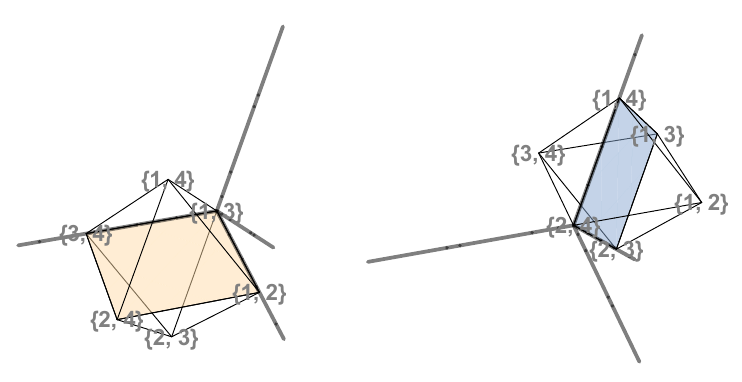}
	\caption{The two (incompatible) arrangements of the blade $((1,2,3,4))$ on the octahedron; these are connected by a square move.  They are incompatible because their superposition does not induce a matroid subdivision: the resulting four tetrahedral cells are not matroid polytopes.  The Steinmann compatibility relation for poles of Feynman diagrams, as reciprocals of linear forms on the kinematic space $\mathcal{K}_{k,n}$ reduces the requirement that the octahedral faces $\simeq\Delta_{2,4}$ of the hypersimplex $\Delta_{k,n}$ be cut into at most two square pyramids.}
	\label{fig:12-7-2019bladearrangementoctahedron}
\end{figure}

\section{Height functions, kinematic space and planar basis}\label{section:height functions}
In this paper, we study a remarkable set of linear forms $\eta_{j_1\cdots j_k}:\mathbb{R}^{\binom{n}{k}}\rightarrow\mathbb{R}$ and in particular their restriction to the \textit{kinematic space}
\begin{eqnarray}\label{eq: kinematic space}
\mathcal{K}_{k,n} = \left\{(s_J)\in\mathbb{R}^{\binom{n}{k}}: \sum_{J\in\binom{\lbrack n\rbrack}{k}:\ J\ni a} s_{J}=0 \text{ for each } a=1,\ldots, n\right\}.
\end{eqnarray}

In geometric terms, the $\eta_J$ correspond to equivalence classes of surfaces over a hypersimplex that are linear over a particular kind of matroid polytope that occurs as the maximal cells in the so-called multi-split matroid subdivisions.

\begin{defn}
	A matroid subdivision is a decomposition $P_1\sqcup\cdots \sqcup P_d$ of a hypersimplex $\Delta_{k,n}$ such that each pair $P_i,P_j$ intersects only on their common facet, and such that each $P_i$ is a matroid polytope.  It moreover a planar (or positroid) subdivision if every maximal cell $\Pi_i$ is a positroid polytope, that is, its facets are given by equations $x_{i}+x_{i+1}+\cdots+x_{i+m}=r_{i,i+m}$ for some integers $r_{i,i+m}$, where $i\in\{1,\ldots, n\}$ and $1\le m \le n-2$, where the indices are assumed to be cyclic. 
\end{defn}

\begin{defn}\label{defn:multisplit}
	Let $d\ge 2$.  A $d$-split of an $m$-dimensional polytope $P$ is a coarsest subdivision $P = P_1\cup\cdots \cup P_d$ into $m$-dimensional polytopes $P_i$, such that the polytopes $P_i$ intersect only on their common facets, and such that 
	$$\text{codim}(P_1\cap\cdots\cap P_d) = d-1.$$  If $d$ is not specified, then we shall use the term multi-split.  
\end{defn}

For any $n$-cycle $\beta = (\beta_1,\ldots, \beta_n)$ define a piecewise-linear function on $\mathbb{R}^n$ by
$$h_\beta(x) = \min\{L_1(x),\ldots, L_n(x)\},$$
where
$$L_j = x_{\beta_{j+1}} + 2x_{\beta_{j+2}}+\cdots (n-1)x_{\beta_{j-1}}.$$
We shall restrict its domain to the hyperplane $\mathcal{H}_{0,n}$ where $x_1+\cdots +x_n=0$.

When $\beta=(1,2,\ldots, n)$ is the standard $n$-cycle, we shall omit $\beta$ and write simply $h(x)$.

\begin{rem}
	The locus of points where the curvature $\nabla^2(h_\beta)$ is nonzero, of a function of the form $h_\beta(x)$, coincides with an object called a blade by A. Ocneanu, see \cite{OcneanuVideo}, where he used the notation $((\beta_1,\beta_2,\ldots, \beta_n))$.  
\end{rem}

In fact the support of the curvature $\nabla^2(h)$ can be expressed using tropical geometry.



\begin{prop}[\cite{Early19WeakSeparationMatroidSubdivision}]\label{prop:blade Tropical hypersurface}
	The blade $\beta = ((1,2,\ldots, n))$ is the tropical hypersurface defined by the bends of the function $h_\beta:\mathcal{H}_{0,n}\rightarrow \mathbb{R}$,
	$$h_{\beta}(x) = \min\{L_1(x),\ldots, L_n(x)\},$$
	that is the locus of points $x$ where the minimum is achieved at least twice.
\end{prop}
\begin{figure}[h!]
	\centering
	\includegraphics[width=1\linewidth]{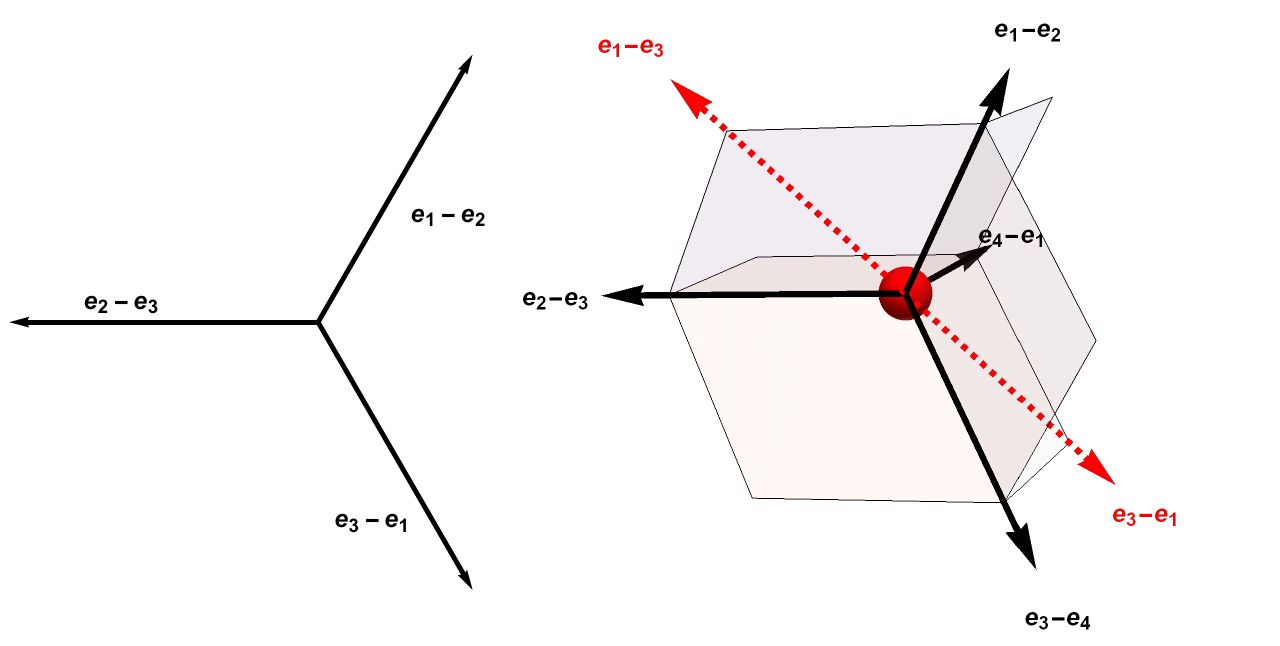}
	\caption{The blades $((1,2,3))$ and $((1,2,3,4))$ are isomorphic to a tropical line (respectively a tropical hyperplane), but they have only cyclic symmetry.  These are used to induce the linear forms $\eta_J$ which appear as poles in the Feynman diagram expansion of the generalized biadjoint amplitude.  See \cite{EarlyBlades,Early19WeakSeparationMatroidSubdivision} for details about blades.}
	\label{fig:tripodandsummedtripods}
\end{figure}
Then $((1,2,\ldots, n))$ is the set where the curvature $\nabla^2(h)$ is nonzero.
Indeed, it is not difficult to see by replacing characteristic functions in \cite{EarlyBlades} by the relevant distributions, that the curvature of the function $h$ expands as a linear combination of products of Heaviside functions and Dirac-Delta functions:
$$\nabla^2(h) = \sum_{1\le a<b\le n}\delta\left(\sum_{j=a}^bx_j\right)\delta\left(\sum_{j=b+1}^{a-1}x_j\right)\prod_{\ell=a}^{b-1}\left(\Theta(x_{a\cdots \ell})\right)\prod_{\ell=b}^{a-1}\left(\Theta(x_{b\cdots \ell})\right)$$
where the indices are cyclic.  For instance,
$$\nabla^2(h_{((1,2,3))}) = \delta(x_1)\delta(x_2+x_3)\Theta(x_2) + \delta(x_2)\delta(x_3+x_1)\Theta(x_3)+\delta(x_3)\delta(x_1+x_2)\Theta(x_1).$$
Translating $h$ to the vertices of hypersimplex $\Delta_{k,n}$ gives rise to a collection of height functions $\rho_J(x) = h(x-e_J)$ for $x\in\Delta_{k,n}$, and restricting these to the vertices of $\Delta_{k,n}$ determines an (integer-valued) height function, which we shall encode by a vector in $\mathbb{Z}^{\binom{n}{k}}$.  First denote by $\rho_J:\Delta_{k,n}\rightarrow \mathbb{R}$ the translation of $h$,
$$\rho_J(x) = h(x-e_J).$$
Now put
$$\mathfrak{h}(e_J):=\sum_{e_I\in\Delta_{k,n}} \rho_J(e_I)e^I \in\mathbb{R}^{\binom{n}{k}}.$$
One can see that the restrictions to $\mathcal{K}_{k,n}$ of the linear functionals dual to the basis $e^J$ are naturally identified with the generalized Mandelstam variables $s_J$.

We now come to one of our main constructions, obtained by dualizing the elements $\mathfrak{h}_{e_J}$, of the planar basis of kinematic invariants.  One can see that the set of these elements $\eta_J$ are invariant under cyclic permutation.

Now we introduce the planar basis.
\begin{defn}
	For any nonfrozen vertex $e_J\in\Delta_{k,n}$, define
	\begin{eqnarray}\label{eq:deltaJ}
		\eta_J(s) = -\frac{1}{n}\sum_{e_I \in\Delta_{k,n}}s_I \rho_J(e_I).
		\end{eqnarray}
\end{defn}

As in \cite{Early19WeakSeparationMatroidSubdivision}, we are are interested in arrangements of the single blade $((1,2,\ldots, n))$ on the vertices of hypersimplices $\Delta_{k,n}$;  Weyl alcoves in $\Delta_{k,n}$ are simplices, \textit{but are not matroid polytopes}, with vertices among those of $\Delta_{k,n}$ and with $n$ facet inequalities of the form $x_{i_1}+x_{i_1+1}+\cdots +x_{i_2}\ge a_{i_{1},i_2}$, where the indices range over a cyclic interval in $\{1,\ldots, n\}$.

Proposition \ref{prop:zero curvature over Weyl alcoves} follows from the observation that the common refinement of all positroid subdivisions is the alcove triangulation of $\Delta_{k,n}$ into $A_{k-1,n-1}$ simplices, where $A_{k,n}$ is the Eulerian number which counts the number of permutations of $\{1,\ldots, n\}$ having $k$ descents.

\begin{prop}\label{prop:zero curvature over Weyl alcoves}
	Any linear combination 
	$$\sum_{e_J\in \Delta_{k,n}} c_J \rho_J(x)$$
	has zero curvature over each Weyl alcove in $\Delta_{k,n}$.
\end{prop}

\begin{prop}The following linear relations hold among linear functionals on respectively $\mathbb{R}^{\binom{n}{k}}$ and $\mathcal{K}_{k,n}$.
	\begin{enumerate}
		\item 	For any frozen vertex $e_J\in\Delta_{k,n}$ where $J = \{j,j+1,\ldots, j+k-1\}$, then the graph of the function $\rho_J(x)$ has constant slope over $\Delta_{k,n}$, hence zero curvature; therefore so does
		$$\sum_{e_I\in\Delta_{k,n}} \rho_J(e_I)e^I \in\mathbb{R}^{\binom{n}{k}}.$$
		Further,	
		$$\eta_J\equiv 0$$
		upon restriction to $\mathcal{K}_{k,n}$.
		\item Given a nonfrozen vertex $e_J\in\Delta_{k,n}$ with $t(\ge 2)$ cyclic intervals, with cyclic initial points say $j_1,\ldots, j_t$, consider the t-dimensional cube 
		$$C_J=\left\{J_L=\{j_1-\ell_1,\ldots, j_t-\ell_t\}: L=(\ell_1,\ldots, \ell_t) \in \{0,1\}^t\right\}.$$
		
		Then the following relation among linear functionals holds identically on $\mathbb{R}^{\binom{n}{k}}$, as well as on the subspace $\mathcal{K}_{k,n}$:
		$$ \sum_{L\in C_J}(-1)^{L\cdot L }\eta_{J_L} = -s_J,$$
		where $L\cdot L$ is the number of 1's in the 0/1 vector $L$.
		\item Moreover, for any frozen vertex $e_J = \sum_{i=0}^{k-1} e_{j+i}$, then
		$$\left(\eta_{j,j+1,\ldots, j+(k-1)} - \eta_{j-1,j+1,\ldots, j+(k-1)}\right)=-s_{J} +\frac{1}{n}\sum_{e_I\in\Delta_{k,n}} s_I,$$
		and upon restriction to $\mathcal{K}_{k,n}$ the second term on the right hand side vanishes and we obtain
		$$ \left(\eta_{j,j+1,\ldots, j+(k-1)} - \eta_{j-1,j+1,\ldots, j+(k-1)}\right)=- \eta_{j-1,j+1,\ldots, j+(k-1)}=-s_{J}.$$
	\end{enumerate}

\end{prop}
\begin{cor}\label{cor: planar basis}
	The set of linear forms $\{\eta_J:e_J\in\Delta_{k,n}\text{ is nonfrozen}\}$ is a basis of functionals on the kinematic space $\mathcal{K}_{k,n}$.  In particular, any pole appearing in the Feynman diagram expansion of $m^{(k)}(\mathbb{I}_n,\mathbb{I}_n)$ can be expanded in the $\eta_J$ basis. 
\end{cor}

Checking compatibility for pairs of planar poles has an efficient implementation via the four-term Steinmann relation, applied to subdivisions induced on the second hypersimplicial faces $\left\{x\in\Delta_{k,n}: x_j=1\text{ for }j\in J\right\}$ as $J$ runs over all $(k-2)$ element subsets of $\{1,\ldots, n\}$.  Thus, the Steinmann relation \cite{Streater75} corresponds combinatorially to a non-crossing condition called weak separation once a cyclic order is fixed.

Checking compatibility for pairs of planar basis elements has a particularly efficient implementation, using weak separation as studied in \cite{Early19WeakSeparationMatroidSubdivision}.  Compatibility for exotic poles which are dual to planar matroid subdivisions which are not multi-splits can also be checked, but the computation (and Mathematica implementation) is somewhat more involved.

\begin{thm}[\cite{Early19WeakSeparationMatroidSubdivision}]\label{thm: weakly separated matroid subdivision blade}
	Given a collection of vertices $e_{I_1},e_{I_2},\ldots, e_{I_m}\in\Delta_{k,n}$, the blade arrangement 
	$$\{((1,2,\ldots, n))_{e_{I_1}},((1,2,\ldots, n))_{e_{I_2}},\ldots, ((1,2,\ldots, n))_{e_{I_m}}\}$$
	induces a matroid (in particular a positroid) subdivision of $\Delta_{k,n}$ if and only if $\{I_1,\ldots, I_m\}$ is weakly separated.	
\end{thm}
Using the facet data from \cite{TropicalGrassmannianCluster Drummond} for $n=8$, we have verified Corollary \ref{cor: generalizedFeynmanDiagramsWeaklySeparated} below in Mathematica, i.e. that every maximal weakly separated collection gives rise to a generalized Feynman diagram in $m^{(3)}(\mathbb{I}_n,\mathbb{I}_n)$, for $n=6,7,8,9$, and we have verified for that every pole of the form $\eta_J$, for $J\in\binom{\lbrack 8\rbrack}{4}$ nonfrozen, i.e., whose indices do not form a single cyclic interval, appears as a pole in the expansion of $m^{(4)}(\mathbb{I}_8,\mathbb{I}_8)$.  We have also computed the expansions of the remaining poles in terms of the $\eta_J$ basis, and using these expansions we have computed the corresponding biadjoint amplitudes.  
\begin{cor}\label{cor: generalizedFeynmanDiagramsWeaklySeparated}
	Every maximal weakly separated collection of (nonfrozen) subsets 
	$$I_1,\ldots, I_{(k-1)(n-k-1)}\in \binom{\lbrack n\rbrack}{k}$$
	defines a generalized Feynman diagram in the sense of \cite{CachazoBorges}:
	$$\prod_{j=1}^{(k-1)(n-k-1)}\frac{1}{\eta_{I_j}}.$$
\end{cor}

Thus, in this sense the sum
\begin{eqnarray}\label{eq: restricted amplitude}
\sum_{\mathcal{C}} \frac{1}{\eta_{J_1}\cdots \eta_{J_{N_k}}},
\end{eqnarray}
over weakly separated collections of $(k-1)(n-(k-1))$ planar basis elements, becomes an inner approximation to the generalized biadjoint scattering amplitude $m^{(k)}(\mathbb{I}_n,\mathbb{I}_n)$.

\begin{example}
	It is straightforward to check relations among linear forms using the straightening relations derived from Equation \eqref{eq: kinematic space}; for instance on $\mathbb{R}^{\binom{6}{3}}$ we have 
	\begin{eqnarray*}
		6\eta_{235} & = & 4 s_{123}+3 s_{124}+2 s_{125}+7 s_{126}+2 s_{134}+s_{135}+6 s_{136}+6 s_{145}+5 s_{146}+4 s_{156}+s_{234}\\
		& + & 5 s_{236}+5 s_{245}+4 s_{246}+3 s_{256}+4 s_{345}+3 s_{346}+2 s_{356}+7 s_{456}\\
		& \equiv  & 6(s_{123}+s_{126}+s_{136}+s_{236})
	\end{eqnarray*}
	and
	\begin{eqnarray*}
		6\eta_{246} & = & 6 s_{123}+5 s_{124}+4 s_{125}+3 s_{126}+4 s_{134}+3 s_{135}+2 s_{136}+2 s_{145}+s_{146}+6 s_{156}+3 s_{234}\\
		& + & 2 s_{235}+s_{236}+s_{245}+5 s_{256}+6 s_{345}+5 s_{346}+4 s_{356}+3 s_{456}\\
		& \equiv & 6(s_{156}+s_{256}+s_{345}+s_{346}+s_{356}+s_{456}),
	\end{eqnarray*}
	upon restriction to $\mathcal{K}_{3,6}$.  Thus, we have $\eta_{246} = R_{12,34,56}$, in the notation of \cite{TropGrassmannianScattering} and elsewhere.
		
	Conversely, 
	\begin{eqnarray*}
		-s_{235} & = & \eta_{235}-\eta_{234}-\eta_{135}+\eta_{134}\\
		-s_{236} & = & \eta_{236}-\eta_{136}-\eta_{235}+\eta _{135}\\
		-s_{246} & = &  \eta _{246}-\eta _{146}-\eta _{236}+\eta _{136}-\eta _{245}+\eta _{145}+\eta _{235}-\eta _{135},
	\end{eqnarray*}
	where we remark that $\eta_{234}\equiv 0$ on $\mathcal{K}_{3,6}$.
	
	Moreover, as a consequence of the relations (``momentum conservation'') on $\mathcal{K}_{3,6}$,
	$$ \eta_{234} - \eta_{134} = -s_{234} + \frac{1}{6}\sum_{e_J\in\Delta_{3,6}} s_J \equiv -s_{234}.$$
\end{example}

\begin{example}\label{example:change of basis matrix}
	One can choose for a basis of linear forms on the (14-dimensional) kinematic space $\mathcal{K}_{3,6} \subset \mathbb{R}^{\binom{6}{3}}$, the set
	$$\{s_{123},s_{124},s_{125},s_{126},s_{134},s_{135},s_{136},s_{145},s_{146},s_{234},s_{235},s_{236},s_{245},s_{246}\}.$$
	The planar basis is 
	$$\{\eta _{124},\eta _{125},\eta _{134},\eta _{135},\eta _{136},\eta _{145},\eta _{146},\eta _{235},\eta _{236},\eta _{245},\eta _{246},\eta _{256},\eta _{346},\eta _{356}\}.$$	
	The change of basis matrix is then
	$$\left(
	\begin{array}{cccccccccccccc}
	0 & 0 & 0 & 0 & 0 & 0 & 0 & 0 & 1 & 0 & 0 & 0 & 0 & 0 \\
	-1 & 0 & 0 & 0 & 0 & 0 & 0 & 0 & -1 & 0 & 1 & 0 & 0 & 0 \\
	1 & -1 & 0 & 0 & 0 & 0 & 0 & 0 & 0 & 0 & -1 & 1 & 0 & 0 \\
	0 & 1 & 0 & 0 & 0 & 0 & 0 & 0 & 0 & 0 & 0 & 0 & 0 & 0 \\
	1 & 0 & -1 & 0 & 0 & 0 & 0 & 0 & 0 & 0 & -1 & 0 & 1 & 0 \\
	-1 & 1 & 1 & -1 & 0 & 0 & 0 & 0 & 0 & 0 & 1 & -1 & -1 & 1 \\
	0 & -1 & 0 & 1 & -1 & 0 & 0 & 0 & 0 & 0 & 0 & 0 & 0 & 0 \\
	0 & 0 & 0 & 1 & 0 & -1 & 0 & 0 & 0 & 0 & 0 & 0 & 0 & -1 \\
	0 & 0 & 0 & -1 & 1 & 1 & -1 & 0 & 0 & 0 & 0 & 0 & 0 & 0 \\
	0 & 0 & 1 & 0 & 0 & 0 & 0 & 0 & 0 & 0 & 0 & 0 & 0 & 0 \\
	0 & 0 & -1 & 1 & 0 & 0 & 0 & -1 & 0 & 0 & 0 & 0 & 0 & 0 \\
	0 & 0 & 0 & -1 & 1 & 0 & 0 & 1 & -1 & 0 & 0 & 0 & 0 & 0 \\
	0 & 0 & 0 & -1 & 0 & 1 & 0 & 1 & 0 & -1 & 0 & 0 & 0 & 0 \\
	0 & 0 & 0 & 1 & -1 & -1 & 1 & -1 & 1 & 1 & -1 & 0 & 0 & 0 \\
	\end{array}
	\right).$$
\end{example}

\section{Boundary operators, generalized Feynman diagrams and Steinmann relations for $\Delta_{2,n-(k-2)}$}\label{sec:boundary operators}
Any subset $L$ of $\{1,\ldots,n\}$ (of size at most $k-2$, say), determines a face of the hypersimplex $\Delta_{k,n}$,
$$\partial_L(\Delta_{k,n}) = \left\{x\in\Delta_{k,n}: x_\ell=1 \text{ for all }\ell\in L\right\},$$
so that in particular whenever $\vert L\vert = k-2$ then $\partial_L(\Delta_{k,n})\simeq \Delta_{2,n-(k-2)}$ modulo affine translation.

We also denote by $\partial_L$ the linear map which is induced on the space of linear combinations of piecewise-continuous functions on $\Delta_{k,n}$ which have zero curvature over every Weyl alcove in $\Delta_{k,n}$, modulo functions having zero curvature over all of $\Delta_{k,n}$.  By Proposition \ref{prop:zero curvature over Weyl alcoves}, the functions $\rho_J$ span a subspace of these.  Indeed, this is a proper subspace, simply because there are only $\binom{n}{k}-n$ nontrivial $\rho_J$'s, while the number of Weyl alcoves in $\Delta_{k,n}$ is the (in general much larger) Eulerian number $A_{k-1,n-1}$.  

Recall the restriction of $h$ to vertices of $\Delta_{k,n}$,
$$\beta_J \mapsto \sum_{e_I\in\Delta_{k,n}} \rho_J(e_I-e_J)e^I$$

In an accompanying Mathematica notebook, the boundary operator $\partial_{\{j\},1}$ introduced in Lemma 24 of \cite{Early19WeakSeparationMatroidSubdivision} is implemented and is used to obtain a criterion for compatibility of coarsest (positroid) subdivisions, and consequently also for compatibility of poles in generalized Feynman diagrams from \cite{CachazoBorges,CachazoPlanarCollections}.

It would be interesting to approach the following conjecture in the context of \cite{Lafforgue} and related work; however it seems somewhat beyond the scope of this paper and we leave it to future work.
\begin{conjecture}
	The set of functions $\rho_J$, where $e_J\in\Delta_{k,n}$ ranges over all $\binom{n}{k}-n$ nonfrozen vertices, define a basis for the space of piecewise-continuous functions which have zero curvature over the maximal cells of some positroid subdivision of $\Delta_{k,n}$.
	
\end{conjecture}

\begin{prop}Specializing to the vertices of $\Delta_{k,n}$ the result holds:
	\begin{enumerate}
	\item When $x$ is in the vertex set of $\Delta_{k,n}$ then we have the identity
	$$\sum_{e_J}h_1(e_J)s_J = \sum_{e_J} a_J \eta_J \text{ and }\sum_{e_J}h_2(e_J)s_J = \sum_{e_J} b_J \eta_J,$$
	where the $s_J$ are coordinate functions and the $\eta_J$ are defined by Equation \eqref{eq:deltaJ}, on the whole space $\mathbb{R}^{\binom{n}{k}}$.  Here the sums are over all $\binom{n}{k}$ vertices of $\Delta_{k,n}$.
	\item When $s_J$ and $\eta_J$ are restricted to the kinematic space $\mathcal{K}_{k,n}\subset\mathbb{R}^{\binom{n}{k}}$, then 
	we again have
	$$\sum_{e_I}h_1(e_I)s_I = \sum_{e_J} a_J \eta_J \text{ and }\sum_{e_I}h_2(e_I)s_I = \sum_{e_J} b_J \eta_J,$$
	but where the sums over $J$ are now over only the \emph{nonfrozen} vertices. 
\end{enumerate}
\end{prop}

Corollary \ref{cor: common refinement criterion} provides the key criterion which we have implemented in the Mathematica notebook which accompanies the arXiv submission: it reduces the problem whether the common refinement of two positroid subdivisions is itself a positroid subdivision to a simple test on the second hypersimplicial faces of $\Delta_{k,n}$.  

\begin{cor}\label{cor: common refinement criterion}
	Suppose $\Pi_1$ and $\Pi_2$ are positroid subdivisions which are induced by (curvatutes of) piecewise continuous functions $h_1,h_2:\Delta_{k,n}\rightarrow\mathbb{R}$.  Then, the common refinement of $\Pi_1$ and $\Pi_2$ is a \emph{positroid} subdivision if and only if the curvatures of $h_1$ and $h_2$ induced on the second hypersimplicial faces $\partial_L(\Delta_{k,n})$ satisfy the Steinmann relations.
\end{cor}
We implement this criterion in an attached Mathematica notebook to compute the maximal cones in the nonnegative tropical Grassmannians $\text{Trop}_+ Gr(k,n)$ for $(k,n)\in \{(3,6),(3,7),(3,8),(3,8),(4,8)\}$.

For the following we need to introduce a simple bijection between two-block planar set partitions $(S,S^c)$ of a set $\{1,\ldots, n\}$ and pairs of integers $(i,j)$.  Namely, we have the following formal identification:
$$\beta_{ij} \leftrightarrow \beta_{((S,S^c))},$$
where, with cyclic indices, we have
$$(S,S^c) = (\{i+1,i+2,\ldots, j\},\{(j+1,j+2\ldots, i)\}).$$

Suppose now that the curvatures $\nabla^2(h_1)$ and $\nabla^2(h_2)$ expand in the basis of planar blades $\{\beta_{e_J}: e_J\text{ is nonfrozen}\},$ as 
$$\nabla^2(h_1) = \sum_{J\in \mathcal{C}_1} a_J \beta_J\text{ and } \nabla^2(h_2) = \sum_{J\in \mathcal{C}_2} b_J \beta_J.$$
The notation $\beta^L_{((S,T))}$ is best understood through an example which follows; see also the Mathematica implementation.

\begin{enumerate}
	\item Compute the boundaries and find the nonzero coefficients $a'_{((S_1,S_2))},b'_{((T_1,T_2))}$ for $$((S_1,S_2))\in\mathcal{D}_1 \text{ and }((T_1,T_2))\in\mathcal{D}_2,$$
	say, such that 
	$$\partial_L\left(\nabla^2(h_1)\right) = \sum_{J\in\mathcal{C}_1}a_J \partial_L(\beta_J) = \sum_{((S_1,S_2))\in\mathcal{D}_1}a'_{((S_1,S_2))}\beta^L_{((S_1,S_2))}$$
	and
	$$\partial_L\left(\nabla^2(h_2)\right) = \sum_{J\in\mathcal{C}_1}b_J \partial_L(\beta_J) = \sum_{((T_1,T_2)) \in\mathcal{D}_2}b'_{((T_1,T_2))}\beta^L_{((T_1,T_2))}.$$
	\item For each pair $((U_1,U_2)),((V_1,V_2)) \in \mathcal{D}_1\cup \mathcal{D}_2$, compute an (affine) analog (which we formulate below) of the Steinmann relation on patterns of intersecting transverse hypersurfaces.  Here our reference for the analogy is the review article \cite{Streater75}, p. 827, to which we refer for details.  \textit{The test yields says that the induced subdivision in the bulk of $\Delta_{k,n}$ is matroidal if at least one of the following intersections is empty:}
	$$U_1\cap V_1\ \ \ U_1\cap V_2$$
	$$U_2\cap V_1\ \ \ U_2\cap V_2.$$
	If all four intersections are nonempty for some such pair $((U_1,U_2)),((V_1,V_2)) \in \mathcal{D}_1\cup \mathcal{D}_2$, then there must be at least one non-matroidal maximal cell in the subdivision of $\Delta_{k,n}$ induced by the superposition of the curvatures $\nabla^2(h_1)$ and $\nabla^2(h_2)$. 
\end{enumerate}

The action of $\partial_{\{j\}}$ on $\beta_{J}$, as well as on $\beta^L_J$, is easily understood by example.  For more details than we can provide here, see Theorem 17 and Lemma 24 and surrounding discussion in \cite{Early19WeakSeparationMatroidSubdivision}.

We shall always sum over faces $j=1,\ldots, n$; therefore let us use the notation
$$\partial = \sum_{j=1}^n\partial_j.$$
Also write 
$$\partial_L = \partial_{\ell_1}\cdots \partial_{\ell_m}$$
if $L=\{\ell_1,\ldots, \ell_m\}$.

We take a (real or complex, but rational numbers $\mathbb{Q}$ suffice for our purposes) graded vector space generated formally by the set
$$\{\beta^L_J: L\in\binom{\lbrack n\rbrack}{m}\text{ and } J\in\binom{\lbrack n\rbrack}{k-m},\text{ for } m=0,\ldots, k-2\}$$ 
Here the grading is on the number of elements in $L$.  The $\beta^L_J$ are subject only to the condition that when the labels of $J$ form a cyclic interval in $\{1,\ldots, n\}\setminus L$ then we declare $\beta^L_J = 0$.

The boundary operator $\partial_j$ acts as follows.  Set $\partial_j(\beta_J) = \beta^{(j)}_{J\setminus\{\ell\}}$, where $\ell=j$ if $j\in J$, and otherwise $\ell$ is the cyclically next element of $\{1,\ldots, n\}$ that is in $J$.  One takes the ``cyclically next element'' in order to match the notation used to encode the subdivision induced on the boundary; in this way our construction is not ad hoc; it is strictly determined geometrically.

This will be more clear with an example.

Let $J = \{1,4,5,6\}$, with $n=8$.  Then $\partial_{1}(\beta_{1456}) = \beta^{(1)}_{456}$.  But the indices $456$ form a cyclic interval, so we declare $\beta^{(1)}_{456} = 0$.  Intuitively this is because $\beta^{(1)}_{456}$, as the curvature of a piecewise continuous function over the hypersimplex $\partial_1(\Delta_{4,8})$, has in fact zero curvature on the interior and consequently induces the trivial subdivision.  But $\partial_2(\beta_{1456}) = \beta^{(2)}_{156}$ is not zero, since $1456$ is not a cyclic interval in $\{1,3,4,5,6,7,8\}$. 

\begin{enumerate}
	\item With $n=6$,
	\begin{eqnarray*}
		\partial(\beta_{134}) & = & \beta^{(1)}_{34} + \beta^{(2)}_{14} + \beta^{(3)}_{14} + \beta^{(4)}_{13} + \beta^{(5)}_{34} + \beta^{(6)}_{34}\\
		& = &  \beta^{(2)}_{14} + \beta^{(3)}_{14} + \beta^{(4)}_{13}\\
		\partial(\beta_{246}) & = & \beta^{(1)}_{46} + \beta^{(2)}_{46} + \beta^{(3)}_{26} + \beta^{(4)}_{26} + \beta^{(5)}_{24} + \beta^{(6)}_{24},
	\end{eqnarray*}
	where in the second line trivial subdivisions have been killed.
	\item It makes sense to extend $\partial$ by linearity:
	\begin{eqnarray*}
		\partial(\beta_{124} + \beta_{346} + \beta_{256} - \beta_{246}) & = & \beta ^{(1)}_{24}+\beta ^{(1)}_{56}+\beta ^{(2)}_{14}+\beta ^{(2)}_{56}+\beta ^{(3)}_{12}+\beta ^{(3)}_{46}+ \beta ^{(4)}_{12}+\beta ^{(4)}_{36}\\
		& + & \beta ^{(5)}_{26}+\beta ^{(5)}_{34}+\beta ^{(6)}_{25}+\beta ^{(6)}_{34}\\
		& = & \beta ^{(1)}_{24}+\beta ^{(2)}_{14}+\beta ^{(3)}_{46}+\beta ^{(4)}_{36} + \beta ^{(5)}_{26}+\beta ^{(6)}_{25},
	\end{eqnarray*}
	modulo trivial subdivisions.
\end{enumerate}
For a more nontrivial example, after some cancellation we have the following for $\Delta_{3,9}$:
\begin{eqnarray*}
	&&\partial\left(-\beta_{258} + \beta_{358} + \beta_{268} + \beta_{259}\right)\\
	& =& \beta^{(1)}_{59}+\beta^{(1)}_{78}+\beta^{(2)}_{59}+\beta^{(2)}_{78}+\beta^{(3)}_{29}+\beta^{(3)}_{58}+\beta^{(4)}_{29}+\beta^{(4)}_{38}+\beta^{(5)}_{29}+\beta^{(5)}_{38}+\beta^{(6)}_{28}+\beta^{(6)}_{35}+\beta^{(7)}_{28}+\beta^{(7)}_{35}\\
	& + & \beta^{(8)}_{27}+\beta^{(8)}_{35}+\beta^{(9)}_{25}+\beta^{(9)}_{78}\\
	& = & \beta^{(1)}_{59}+\beta^{(2)}_{59}+\beta^{(3)}_{29}+\beta^{(3)}_{58}+\beta^{(4)}_{29}+\beta^{(4)}_{38}+\beta^{(5)}_{29}+\beta^{(5)}_{38}+\beta^{(6)}_{28}+\beta^{(6)}_{35}+\beta^{(7)}_{28}+\beta^{(7)}_{35}\\
	& + & \beta^{(8)}_{27}+\beta^{(8)}_{35}+\beta^{(9)}_{25}.
\end{eqnarray*}

Note that when $\beta$'s with the same superscript are collected, the subscripts define weakly separated collections.  For instance,
$$\beta^{(3)}_{29}+\beta^{(3)}_{58}$$
corresponds to the planar subdivision of $\partial_{\{3\}}(\Delta_{3,9}) = \left\{x\in\Delta_{3,9}: x_3=1\right\}\simeq \Delta_{2,8}$ induced by the affine hyperplanes
$$\left\{x\in \Delta_{3,9}: x_3=1,\ x_{12}=1 = x_{456789} \right\} \text{ and } \left\{x\in \Delta_{3,9}: x_3=1,\ x_{12459}=1 = x_{678} \right\}.$$
Then it is easy to see that the pair $(S,S^c) = (12,456789)$ and $(T,T^c) = (12459,678)$ satisfy the Steinmann relations from Corollary \ref{cor: common refinement criterion}.  Now repeat for each facet $\partial_{\{j\}}(\Delta_{3,9})$.

This means geometrically that the curvature of the function 
$$-h_{258} + h_{358} + h_{268} + h_{259}$$
induces on each second hypersimplicial face of $\Delta_{3,9}$ a positroid subdivision, from which it follows that a positroid subdivision is induced in the bulk of $\Delta_{3,9}$!

Finally, let us compute one simple example for $\Delta_{4,8}$, as can be easily implemented in Mathematica with the command
$$\frac{1}{2}\text{bb}\lbrack\{1,2,4,5\}\rbrack \slash \slash \text{pd} \slash \slash \text{pd}\slash\slash \text{modTrivSubd}.$$
This gives
\begin{eqnarray*}
	\sum_{1\le i<j\le n}\partial_i\partial_j\left(\beta_{1245}\right) & = & \beta^{(13)}_{25}+\beta^{(14)}_{25}+\beta^{(15)}_{24}+\beta^{(23)}_{15}+\beta^{(24)}_{15}+\beta^{(25)}_{14}+\beta^{(36)}_{25}+\beta^{(46)}_{25}+\beta^{(56)}_{24},
\end{eqnarray*}
where we have killed $\beta^{a,b}_J$'s which correspond to identically zero curvatures on the corresponding face $\partial_{a,b}(\Delta_{4,8})$.  The terms sent to zero consist of those $\beta^{a,b}_{ij}$ such that $i$ and $j$ are cyclically adjacent in the set $\{1,\ldots, 8\}\setminus\{a,b\}$, with respect of course to the standard cyclic order $(12\cdots 8)$.  With direct translation, this sum becomes an array of (degenerate) trees, as in \cite{CachazoPlanarCollections}.

\section{Additional applications}
The purpose of following two sections is to preview \cite{Early2020} with some additional applications of the techniques developed in \cite{Early19WeakSeparationMatroidSubdivision} and in this work.

\subsection{Amplitude condensation examples: $\mathcal{K}_{3,6}$ and $\mathcal{K}_{3,7}$}

\begin{example}
	
	It follows from \cite{TropGrassmannianScattering} that there are eight Laurent monomials in $m^{(3)}(\mathbb{I}_6,\mathbb{I}_6)$ which contain the element $\eta_{246}$: these 8 out of 34 maximal weakly separated collections containing (the reciprocal of) $\eta_{246}$ combine to give
	\begin{eqnarray}\label{eq:3-split X36}
	\sum_{\{\eta_{J_1},\eta_{J_2},\eta_{J_3},\eta_{J_4}\}\ni\eta_{246}}\frac{1}{\eta_{J_1}\eta_{J_2}\eta_{J_3}\eta_{J_4}} & = & \left(\frac{1}{\eta_{124}}+\frac{1}{\eta_{236}}\right)\left(\frac{1}{\eta_{146}}+\frac{1}{\eta_{256}}\right)\left(\frac{1}{\eta_{245}}+\frac{1}{\eta_{346}}\right)\frac{1}{\eta_{246}}\nonumber\\
	& = & \frac{\left(\eta_{124}+\eta_{236}\right) \left(\eta_{146}+\eta_{256}\right) \left(\eta_{245}+\eta_{346}\right)}{\eta_{124} \eta_{146} \eta_{236} \eta_{245}  \eta_{256} \eta_{346}}\frac{1}{\eta_{246}}. 
	\end{eqnarray}
	Together its cyclic twist $\eta_{135}$, we have accounted for 16 out of the 34 maximal weakly separated collections.  In the notation of \cite{TropGrassmannianScattering}, Equation \eqref{eq:3-split X36} becomes
	$$\sum_{\{\eta_{J_1},\eta_{J_2},\eta_{J_3},\eta_{J_4}\}\ni\eta_{246}}\frac{1}{\eta_{J_1}\eta_{J_2}\eta_{J_3}\eta_{J_4}} = \left(\frac{1}{t_{1256}}+\frac{1}{s_{123}}\right)\left(\frac{1}{s_{156}}+\frac{1}{t_{3456}}\right)\left(\frac{1}{s_{345}}+\frac{1}{t_{1234}}\right)\frac{1}{R_{12,34,56}}.$$
	
	These can be used to simplify the amplitude; we focus for now on the part of the sum which contains only the 14 basis elements $\eta_J$.  Consider the following Laurent polynomial on $\mathcal{K}_{3,6}$:
	\begin{eqnarray*}
		& & \frac{\left(\eta_{124}+\eta_{236}\right) \left(\eta_{146}+\eta_{256}\right) \left(\eta_{245}+\eta_{346}\right)}{\eta_{124} \eta_{146} \eta_{236} \eta_{245} \eta_{256} \eta_{346}}\frac{1}{\eta_{246} } + \frac{\left(\eta_{125}+\eta_{136}\right) \left(\eta_{134}+\eta_{235}\right) \left(\eta_{145}+\eta_{356}\right)}{\eta_{125} \eta_{134} \eta_{136} \eta_{145} \eta_{235} \eta_{356}}\frac{1}{\eta_{135} }\\
		& + & \sum_{\{J_1,J_2,J_3,J_4\}:\text{ 2-splits only}} \frac{1}{\eta_{J_1}\eta_{J_2}\eta_{J_3}\eta_{J_4}}.
	\end{eqnarray*}
	As the second line is a sum of over all 18 maximal weakly separated collections we have accounted for all $34 = (8+8)+18$ pole configurations in the restricted amplitude of Equation \eqref{eq: restricted amplitude}.  The amplitude $m^{(3)}(\mathbb{I}_6,\mathbb{I}_6)$ contains 16 more; these are obtained by including the linear forms dual to the 3-splits that are induced by the blades respectively $ ((12_1 56_1 34_1)) \text{ and } ((16_1 45_1 23_1)).$
	Thus we have decomposed the amplitude $m^{(3)}(\mathbb{I}_6,\mathbb{I}_6)$ into five distinct groups: four kinds of arrangements containing exactly one 3-split, and one big group consisting of only 2-splits.
\end{example}

\begin{example}
	Similarly, summing over for instance the set of all 12 maximal weakly separated collections of vertices of $\Delta_{3,7}$ containing containing both $e_{247}$ and $e_{257}$ gives
\begin{eqnarray}\label{eqn:all collections containing two 2splits D37}
\sum_{\{\eta_{J_j}\}\ni\eta_{247},\eta_{257}} \prod_{j=1}^6\frac{1}{\eta_{J_j}}
& = & \left(\frac{1}{\eta _{237}}+\frac{1}{\eta _{124}}\right) \left(\frac{1}{\eta _{267}}+\frac{1}{\eta _{157}}\right) \left(\frac{1}{\eta _{245} \eta _{256}}+\frac{1}{\eta _{245} \eta _{457}}+\frac{1}{\eta _{347} \eta _{457}}\right)\nonumber\\
& \times & \left(\frac{1}{\eta_{247}\eta_{257}}\right).
\end{eqnarray}

There is a single cyclic permutation class of exotic coarsest positroid subdivisions for $\Delta_{3,7}$.  One of its representatives is (dual to) the form
$$W:=s_{123}+s_{124}+s_{134}+s_{234}+s_{345}+s_{346}+s_{347}+s_{356}+s_{456}=\eta _{247}+\eta _{256}+\eta _{346}-\eta _{246},$$
as discussed in \cite{CachazoRojas}.

Let us now use the data from \cite{TropicalGrassmannianCluster Drummond} to condense the amplitude around all generalized Feynman diagrams in the $X(3,7)$ amplitude containing the 3-splits $\eta_{247}$ and $\eta_{257}$.  Then we find 

\begin{eqnarray}\label{eq:X37 associahedron 2d}
&&\left(\frac{1}{\eta _{267}}+\frac{1}{\eta _{157}}\right) \left(\frac{1}{\eta _{237}}+\frac{1}{\eta _{124}}\right) \left(\frac{1}{\eta _{245} \eta _{256}}+\frac{1}{\eta _{245} \eta _{457}}+\frac{1}{\eta _{347} \eta _{457}}+\frac{1}{\eta _{347} W}+\frac{1}{\eta _{256} W}\right)\nonumber\\
&&\times \left(\frac{1}{\eta _{247} \eta _{257}}\right).
\end{eqnarray}
On the other hand, if we simply neglect $W$ then clearly what remains is Equation \eqref{eqn:all collections containing two 2splits D37}.  See Figure \ref{fig:12-17-2019fiberassociahedron2d}.
\begin{figure}[h!]
	\centering
	\includegraphics[width=0.7\linewidth]{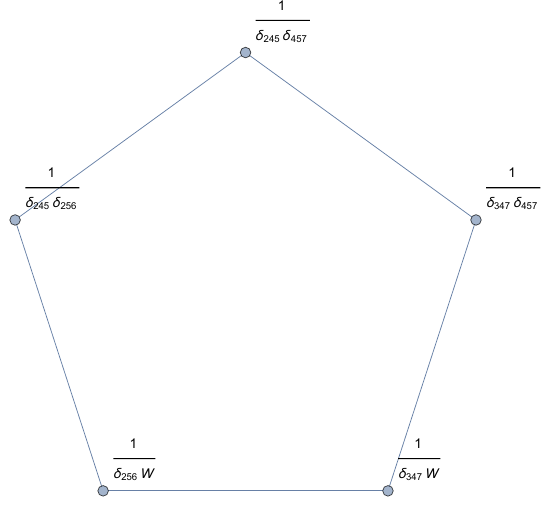}
	\caption{When the exotic pole $W$ is included, the third factor in Equation \eqref{eq:X37 associahedron 2d} is completed combinatorially to the 2-d associahedron.}
	\label{fig:12-17-2019fiberassociahedron2d}
\end{figure}

Now, on the other hand condensing around this particular $W$ we find 
\begin{eqnarray*}
	& & \left(\frac{1}{\eta _{256}}+\frac{1}{\eta _{347}}\right)\left(\frac{1}{\eta _{257} \eta _{267}}+\frac{1}{\eta _{267} \eta _{346}}+\frac{1}{\eta _{157} \eta _{257}}+\frac{1}{\eta _{157} R_{17,56,234}}+\frac{1}{\eta _{346} R_{17,56,234}}\right)\\
	& \times & \left(\frac{1}{\eta _{237} \eta _{247}}+\frac{1}{\eta _{237} \eta _{356}}+\frac{1}{\eta _{124} \eta _{247}}+\frac{1}{\eta _{124} R_{127,56,34}}+\frac{1}{\eta _{356} R_{127,56,34}}\right)\left(\frac{1}{W}\right).\end{eqnarray*}

Here $R_{17,56,234}$ is the linear form dual to the blade $((17_1 56_1 234_1))$; one can check that it has the expansion
$$R_{17,56,234} = \eta_{156}+\eta_{346} + \eta_{147}- \eta_{146}.$$
\end{example}

\begin{figure}[h!]
	\centering
	\includegraphics[width=0.7\linewidth]{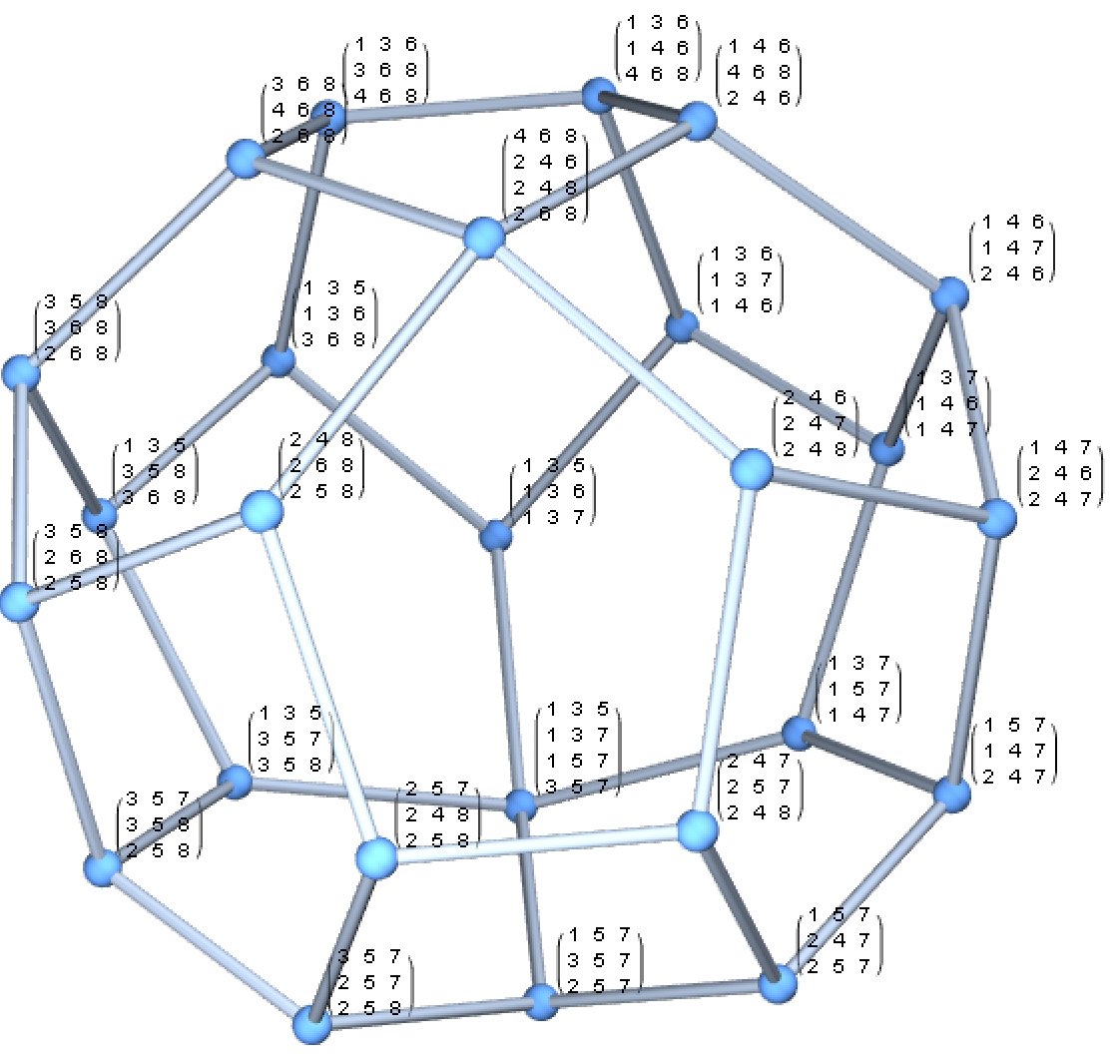}
	\caption{The exchange graph for 3-splits on $\Delta_{3,8}$ corresponding to planar basis elements $\eta_{abc}$ with none of $a,b,c$ cyclically adjacent.  Vertices are thus labeled by weakly separated collections of these.}
	\label{fig:12-9-2019threesplitcollections}
\end{figure}
\subsection{Straightening the potential function and generalized cross-ratios}
In the usual formulation for $n$ points on the Riemann sphere, one has a potential function $\mathcal{P}_{2,n}:Gr(2,n)\times \mathcal{K}_{2,n}\rightarrow \mathbb{C}$ by 
$$\mathcal{P}_{2,n} = \sum_{1\le i<j\le n} s_{ij} \log(d_{ij}),$$
where $d_{ij}$ is the $2\times 2$ minor of a $2\times n$ matrix, with the columns indexed by $\{i,j\}$, and $s_{ij}$ are the coordinate functions, or Mandelstam invariants, on $\mathcal{K}_{2,n}$.  The Mandelstam invariants satisfy the $n$ linearly independent relations
$$\sum_{j\not=i} s_{ij}=0$$
which make the potential function well-defined on the quotient of $Gr(2,n)$ by the torus $(\mathbb{C}^\star)^n$.

Denote by $d_{J}$ the determinant of the $k \times k$ submatrix with column set $J$ of a $k\times n$ matrix. 

We shall diagonalize the potential function on $\mathcal{K}_{k,n}$ using the planar basis of linear forms $\eta_J$.

Now whenever $2\le k\le n-2$, one has a generalized potential function $\mathcal{P}_{k,n}: Gr(k,n) \times\mathcal{K}_{k,n}\rightarrow
\mathbb{C}$, defined by 
$$\mathcal{P}_{k,n} = \sum_{e_J \in\Delta_{k,n}} s_J \log(d_J),$$
where one sums over all $k$-element subsets of $\{1,\ldots, n\}$.

 For convenience, we record the first few cases of the potential function.

In the planar basis, 
$$\mathcal{P}_{24} = \eta _{13} \log \left(\frac{d_{14} d_{23}}{d_{13} d_{24}}\right)+\eta _{24} \log \left(\frac{d_{12} d_{34}}{d_{13} d_{24}}\right),$$
and
$$\mathcal{P}_{25} = \eta _{13} \log \left(\frac{d_{14} d_{23}}{d_{13} d_{24}}\right)+\eta _{14} \log \left(\frac{d_{15} d_{24}}{d_{14} d_{25}}\right)+\eta _{24} \log \left(\frac{d_{25} d_{34}}{d_{24} d_{35}}\right)+\eta _{25} \log \left(\frac{d_{12} d_{35}}{d_{13} d_{25}}\right)+\eta _{35} \log \left(\frac{d_{13} d_{45}}{d_{14} d_{35}}\right).$$
Denoting by $u_{i+1,j+1} := \frac{d_{i+1,j}d_{i,j+1}}{d_{i,j}d_{i+1,j+1}}$ the coefficient of $\eta_{ij}$, then $u_{i,j}$ coincides with Equation 6.7 of \cite{2018Worldsheet}.  Matroidal blade arrangements guide us to the natural analog for $k\ge 3$, where we start to have higher order cross-ratios.

By straightening to the planar basis of linear forms $\eta_J$, for $\mathcal{P}_{3,6}$ one derives
\begin{eqnarray*}
	\mathcal{P}_{3,6} & = & \eta _{124} \log \left(\frac{d_{125} d_{134}}{d_{124} d_{135}}\right)+\eta _{125} \log \left(\frac{d_{126} d_{135}}{d_{125} d_{136}}\right)+\eta _{134} \log \left(\frac{d_{135} d_{234}}{d_{134} d_{235}}\right)+ \eta _{146} \log \left(\frac{d_{156} d_{246}}{d_{146} d_{256}}\right)\nonumber\\
	& + &\eta _{136} \log \left(\frac{d_{146} d_{236}}{d_{136} d_{246}}\right)+\eta _{145} \log \left(\frac{d_{146} d_{245}}{d_{145} d_{246}}\right)+\eta _{235} \log \left(\frac{d_{236} d_{245}}{d_{235} d_{246}}\right)+\eta _{236} \log \left(\frac{d_{123} d_{246}}{d_{124} d_{236}}\right)\nonumber\\
	& + &\eta _{245} \log \left(\frac{d_{246} d_{345}}{d_{245} d_{346}}\right)+\eta _{256} \log \left(\frac{d_{125} d_{356}}{d_{135} d_{256}}\right)+\eta _{346} \log \left(\frac{d_{134} d_{356}}{d_{135} d_{346}}\right)+\eta _{356} \log \left(\frac{d_{135} d_{456}}{d_{145} d_{356}}\right)\\
	& + & \eta _{135} \log \left(\frac{d_{136} d_{145} d_{235} d_{246}}{d_{135} d_{146} d_{236} d_{245}}\right)+ \eta _{246} \log \left(\frac{d_{124} d_{135} d_{256} d_{346}}{d_{125} d_{134} d_{246} d_{356}}\right).\nonumber
\end{eqnarray*}

We now give the general expression for the coefficients which appear by straightening the potential function to the planar basis, by assigning to each nonfrozen vertex $e_J$ of the hypersimplex a generalized cross-ratio, with minors labeled by the vertices of a $t$-dimensional cube, where $t$ is the number of cyclic intervals in the set $J$.

So given a nonfrozen vertex $e_J\in\Delta_{k,n}$ with $t(\ge 2)$ cyclic intervals, with cyclic endpoints say $j_1,\ldots, j_t$, consider the t-dimensional cube 
$$C_J=\left\{J_L=\{j_1+\ell_1,\ldots, j_t+\ell_t\}: L=(\ell_1,\ldots, \ell_t) \in \{0,1\}^t\right\}.$$
With  
$$\text{Num}_J = \left\{J_L: \sum_{i\in L}i\ \text{ is odd}\right\}\ \ \text{ and } \ \ \text{Den}_J = \left\{J_L: \sum_{i\in L} i\ \text{ is even}\right\},$$
define 
$$w_J = \frac{\prod_{M\in \text{Num}_J} d_M}{\prod_{M\in \text{Den}_J} d_M}.$$
\begin{rem}
	Unfortunately these generalized cross-ratios $w_J$ can not in any obvious way be used to define coordinate systems to extract face data in the same way that is possible for the cross-ratio coordinates\footnote{We thank N. Arkani-Hamed for this observation.} $u_{ij}$, as was very recently done in \cite{NimaHeLamThomas} for so-called binary geometries.  However, it seems like a natural question to ask what happens to the space configuration space $X(k,n)$ of $n$ points in $\mathbb{CP}^{k-1}$, modulo $GL(k)$ when they the $w_J$ are required to be (1) real and (2) in the interval $\lbrack 0,1\rbrack$.  Clearly the (torus quotient of the) nonnegative Grassmannian satisfies this property, but our proposal is potentially more general.  We leave such questions to future work.
	
\end{rem}
\begin{prop}\label{prop:straightenedPotential}
	After straightening the potential function to the planar basis of linear forms on $\mathcal{K}_{k,n}$, the coefficient of $\eta_J$ is (the logarithm of) the generalized cross-ratio $w_J$.  We have
	$$\mathcal{P}_{k,n} = \sum_{e_J\in\Delta_{k,n}\text{ nonfrozen}} \eta_J \log(w_J).$$
\end{prop}

The product of all generalized cross-ratios has a simple form.

\begin{prop}\label{example: product all planar cross ratios}
	Taking the product over all $\binom{n}{k}-n$ generalized cross-ratios we get
	$$\prod_{e_J\in\Delta_{k,n}\text{ nonfrozen}}w_J = \frac{\prod_{j=1}^n d_{I_j}}{\prod_{j=1}^n d_{L_j}},$$
	where $d_{I_j} = d_{j,j+1,\ldots, j+k-1}$ and $d_{L_j} = d_{j,j+1,\ldots, j+k-2,j+k}$, where the indices are cyclic.  
\end{prop}

For instance, for $X(3,6)$ we have
$$\prod_{e_J\in\Delta_{3,6}\text{ nonfrozen}}w_J =\frac{d_{123} d_{234} d_{345} d_{456} d_{561} d_{612}}{d_{124} d_{235} d_{346} d_{451} d_{562} d_{613}}.$$

\section{Discussion}\label{sec:discussion}
Evidently much work with the $\eta_J$ basis remains; in this work we have only sketched an outline of what is possible.  Let us now discuss some possible directions.

First, we remark that while the expansion of the full amplitude biadjoint amplitude $m^{(k)}(\mathbb{I}_n,\mathbb{I}_n)$ is for now inaccessible when one restricts to generalized Feynman diagrams involving only poles of the form $\eta_J$, doing so has at least one key advantage: first, the data for the part accessible with only $\eta_J$'s is extremely compact and one can generate pole data to larger $k$ and $n$ than if the whole amplitude were included.  The poles, as linear forms $\eta_J$, are constructed uniquely using Equation \eqref{eq:deltaJ} from a single $k$-element set, and generalized Feynman diagrams are $(k-1)(n-k-1)$-element collections of these.  The compatibility rule for planar basis poles is computationally extremely efficient: to check whether $\eta_{J_1}$ and $\eta_{J_2}$ are compatible, simply compute the difference $e_{J_1}-e_{J_2}$ and look for the pattern $(1,-1,1,-1)$ or $(-1,1,-1,1)$, where $e_J = \sum_{j\in J} e_j$.  This is simply a restatement of the weak separation condition.  For example, $\eta_{147}$ and $\eta_{257}$ are incompatible because the difference
$$e_{147} - e_{257} = (1,-1,0,1,-1,0,\ldots).$$
contains the bad pattern.  Consequently one can very efficiently generate, store and analyze a nontrivial chunk of the amplitude for large $k$ and $n$.  This missing portion consists of all maximal positroid subdivisions which contain at least one exotic pole, i.e. those coarsest positroid subdivisions that are not multi-splits of the form $\eta_J$, see Definition \ref{defn:multisplit}.  One question is to study the expansions of the exotic poles in the $\eta_J$ basis and find a scheme to derive and classify the exotic poles.  Such issues are left to future work \cite{Early2020}.

Let us recall Section \ref{sec:boundary operators}, where we drew an analogy between the Steinmann relations, formulated in \cite{Streater75} as a set of linear equations on discontinuities of generalized retarded functions, and a combinatorial condition on pairs of affine hyperplanes which intersect the interior of the second hypersimplex $\Delta_{2,n}$.  We simply remark that in the context of this work, it seems natural to wonder about the possibility to ``extend'' the original Steinmann relations into the interiors of the hypersimplices $\Delta_{k,n}$ for $k\ge 3$.

For planar $\mathcal{N}=4$ SYM amplitudes, as discussed very recently in \cite{NimaLamSpradlin2019}, one would have to develop a thorough and constructive understanding of the positroid subdivisions of $\Delta_{4,n}$ for $n$ beyond $8$ or $9$.  The problem is that the totally positive tropical Grassmannians one obtains for the amplitude are high in dimension and complexity, making direct visualization impossible, but there are some clues about possible ways around this.  Indeed, see Figure \ref{fig:12-12-2019d48condensationgraph2d} for the complex built by refining those coarsest positroid subdivisions of $\Delta_{4,8}$ that correspond to planar poles $\eta_J$ where $J$ has either three or four cyclic intervals; one can see (using say GraphPlot3D in Mathematica, as we have done) that the adjacency graph embeds nicely into $\mathbb{R}^3$ as the 1-skeleton of a polyhedron (though there are two problematic vertices at the north and south poles, perhaps similarly to what was found in \cite{GalashinPostnikovWilliams} for $(3,6)$).  Now Figure \ref{fig:foursplitsg410} shows that for positroid subdivisions of $\Delta_{4,10}$, if one keeps only 4-splits, i.e. $\eta_{abcd}$ with $abcd$ not cyclically adjacent, treating 2-splits and 3-splits as ``coefficients,'' then the adjacency graph for the generalized Feynman diagrams is the one-skeleton of a polyhedral complex that \textit{still } embeds nicely in dimension three!  In contrast, the ambient dimension of the relevant object, the (nonnegative) tropical Grassmannian $\text{Trop}(4,10)$, is significantly more: the maximal faces have dimension $21=(k-1)(n-k-1) $.  Finally, it would be very interesting to study any special kinematics arising from the planar basis; for instance, in the case $k=2$, setting the $\eta_J$ to $1$ corresponds to taking the special ``Catalan'' kinematics from \cite{CachazoHeYuan}.

\textbf{Note added.}  While preparing this paper for submission we learned of the work \cite{HeRenZhang}, which has some overlap with ours.

\acknowledgments
We thank Nima Arkani-Hamed, Freddy Cachazo, Alfredo Guevara, Thomas Lam, Jeanne Scott, Bruno Umbert and Yong Zhang for useful comments and discussion.  This research was supported in part by Perimeter Institute for Theoretical Physics. Research at Perimeter Institute is supported by the Government  of  Canada  through the Department of Innovation, Science and Economic Development Canada and by the Province of Ontario through the Ministry of Research, Innovation and Science.

\appendix
\section{Enumeration of maximal weakly separated collections}

In \cite{Early19WeakSeparationMatroidSubdivision} we enumerated the maximal weakly separated collections, (that is, the number of maximal matroidal blade arrangements on $\Delta_{k,n}$) in Mathematica with the help of the FindClique algorithm; the counts are given in the table below, for rows with $n=4,5,\ldots, 12$ and columns with $k=2,3,\ldots, n-2$.  Note that these give a (very) lower bound for the number of generalized Feynman diagrams when exotic poles are included; for instance, the number for $m^{(4)}(\mathbb{I}_{11},\mathbb{I}_{11})$ will be significantly larger than the 71 million finest subdivisions involving only multi-splits in the table below.  

\begin{small}
	$$\begin{tabular}{c|cccccccccc}
$n\setminus k$ &2&3&4&5&6&7&8&9& 10\\
\hline 
4 & 2 &  &  &  &  &  &  & & \\
5  & 5 & 5 &  &  &  &  &  &  &\\
6 & 14 & 34 & 14 &  &  &  &  &  & \\
7 & 42 & 259 & 259 & 42 &  &  &  &  & \\
8 & 132 & 2136 & 5470 & 2136 & 132 &  &  & &  \\
9 & 429 & 18600 & 122361 & 122361 & 18600 & 429 & & &  \\
10 & 1430 & 168565 & 2889186 & 7589732 & 2889186 & 168565 & 1430 & & \\
11 & 4862 & 1574298 &71084299 & & &71084299 & 1574298 & 4862 & \\
12 & 16796 & 15051702 & & & & & & 15051702 & 16796
\end{tabular} $$
\end{small}

\section{Enumeration of generalized Feynman diagrams by degree of denominator}\label{sec: Feynman Diagram Enumeration}

The table below counts the number of finest positroid subdivisions of $\Delta_{3,8}$ by explicitly tabulating maximal collections of compatible coarsest planar subdivisions, using the ray data from \cite{TropicalGrassmannianCluster Drummond}.  

For our dataset we use the 120 tropical Plucker vectors which were shown in \cite{TropicalGrassmannianCluster Drummond} to be generating rays of the nonnegative tropical Grassmannian $\text{Trop}(G(3,8))$.  Each of these defines a linear functional on the kinematic space, that is a sum of generalized Mandelstam variables.  
\begin{enumerate}
	\item Each tropical Plucker vector is dual to a linear functional on the kinematic space, i.e. it is a sum of generalized Mandelstam invariants.
	\item For each tropical Plucker vector, expand its linear functional in the planar basis of $\eta_J$'s.
	\item Replace each $\eta_J$ with the curvature $\beta_J$.
	\item Now given any pair of tropical Plucker vectors $\pi,\pi'$, perform steps (1), (2), and (3) to each, apply the boundary operator $$\sum_{L\in\binom{\lbrack n\rbrack}{n-(k-2)}}\partial_{(L,1)}$$ 
	and then check the Steinmann relations for the linear combination of blades on each second hypersimplicial face $\partial_{L}(\Delta_{k,n})$ of $\Delta_{k,n}$, where $L\in\binom{\lbrack n\rbrack}{k}$.  
\end{enumerate}
In summary:
\begin{eqnarray*}
\pi & = & \sum_{J}c_Je^J \mapsto \sum_{J}c_J s_J = \sum_{J\text{ nonfrozen}}b_J \eta_J\mapsto  \sum_{J\text{ nonfrozen}}b_J \beta_J\\
& \mapsto  & \sum_{L\in\binom{\lbrack n\rbrack}{n-(k-2)}}\partial_{(L,1)}\left(\sum_{J\text{ nonfrozen}}b_J \beta_J\right)
\end{eqnarray*}
and similarly for $\pi'$, and finally check whether the Steinmann relations hold on each second hypersimplicial face $\partial_{L}(\Delta_{k,n})\simeq\Delta_{2,n-(k-2)}$ among the affine hyperplanes corresponding to the curvatures $\beta^{(L)}_{ab}$.

Using the 120 rays from the data given in \cite{TropicalGrassmannianCluster Drummond} for $\text{Trop}Gr_+(3,8)$ one finds 13612 maximal dimension cones, in agreement with \cite{CachazoPlanarCollections}.  For the amplitude $m^{(3)}(\mathbb{I}_8,\mathbb{I}_8)$ this corresponds to counting the number $g$ of generalized Feynman diagrams such that the denominator of the corresponding rational function has degree $d$.  The breakdown is
$$
\begin{tabular}{|c|c|c|c|c|c|c|c|c|c|c|c|}
\hline 
d & 8 & 9 & 10 & 11 & 12 & 13 & 14 & 15 & 16 & 17 \\
\hline 
g & 9672 & 1696 & 1092 & 480 & 416 & 104 & 88 & 32 & 24 & 8 \\
\hline 
\end{tabular}
$$

For $m^{(3)}(\mathbb{I}_9,\mathbb{I}_9)$, using a given set\footnote{We thank Yong Zhang for sharing his datasets of rays for $\text{Trop}Gr_+(3,9)$ and $\text{Trop}Gr_+(4,8)$.} of rays for $\text{Trop} Gr_+(3,9)$, we computed the 346710 maximal dimension cones.  They have the following breakdown:
\begin{eqnarray*}
	&& \begin{tabular}{|c|c|c|c|c|c|c|c|c|c|c|c|c|}
		\hline 
		d& 10 & 11 & 12 & 13 & 14 & 15 & 16 & 17 & 18 & 19 & 20 & 21 \\ 
		\hline 
		g&  186147 & 46395 & 35181 & 19854 & 20472 & 9666 & 9171 & 4929 & 4188 & 2817 & 2415 & 1203 \\ 
		\hline 
	\end{tabular}\\
	&&\begin{tabular}{|c|c|c|c|c|c|c|c|c|c|c|c|c|c|c|c|c|c|}
		\hline 
		d& 22 & 23 & 24 & 25 & 26 & 27 & 28 & 29 & 30 & 31 & 32 & 33 & 35 & 36 & 37 & 39 & 46 \\ 
		\hline 
		g&  1314 & 774 & 666 & 396 & 288 & 240 & 129 & 117 & 153 & 75 & 18 & 18 & 36 & 18 & 6 & 18 & 6 \\ 
		\hline 
	\end{tabular}
\end{eqnarray*}

Similarly, from the 360 rays for $\text{Trop}_+Gr(4,8)$ we find 90608 maximal dimensional cones again in agreement with \cite{CachazoPlanarCollections}; the number $g$ of generalized Feynman diagrams having a degree $d$ denominator has the following breakdown.
\begin{eqnarray*}
	&& \begin{tabular}{|c|c|c|c|c|c|c|c|c|c|c|c|c|}
		\hline 
		d& 9 & 10 & 11 & 12 & 13 & 14 & 15 & 16 & 17 & 18 & 19 & 20\\ 
		\hline 
		g&  50356 & 12320 & 9116 & 6064 & 4448 & 2332 & 2176 & 872 & 976 & 676 & 384 & 336\\ 
		\hline 
	\end{tabular}\\
&&\begin{tabular}{|c|c|c|c|c|c|c|c|c|c|c|c|c|}
	\hline 
	d& 21 & 22 & 23 & 24 & 25 & 26 & 27 & 29 & 33 & 34 & 36 & 49 \\ 
	\hline 
	g&  200 & 48 & 8 & 80 & 72 & 24 & 48 & 16 & 20 & 16 & 16 & 4 \\ 
	\hline 
\end{tabular} 
\end{eqnarray*}

\newpage

\begin{figure}[h!]
	\centering
	\includegraphics[angle=90,origin=c,width=1.05\linewidth]{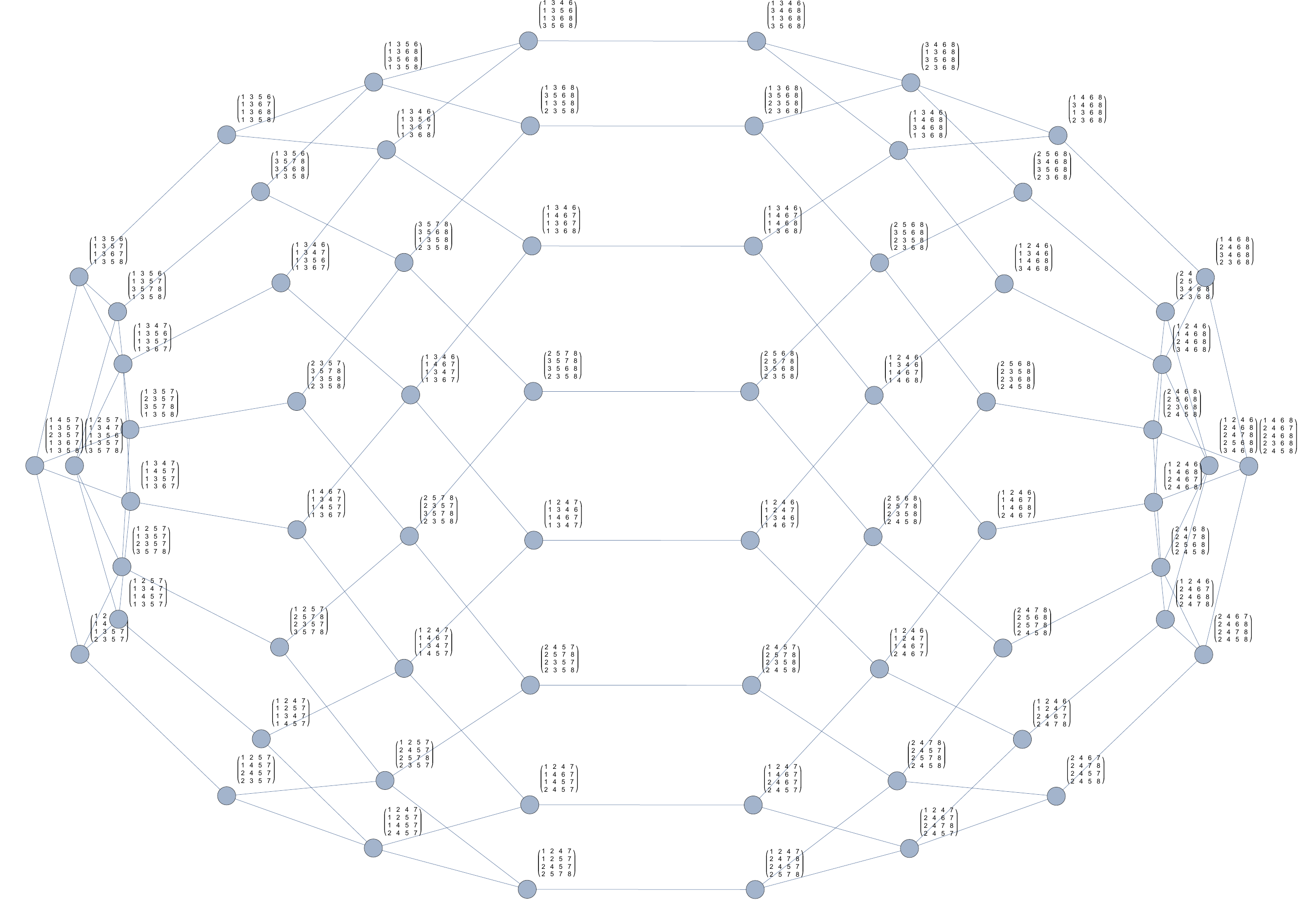}
	\caption{Condensation of the multi-split complex of $\eta_J$'s for $\Delta_{4,8}$ around 3-splits and 4-splits.}
		\label{fig:12-12-2019d48condensationgraph2d}
\end{figure}
\begin{figure}[h!]
	\centering
	\includegraphics[angle=90,origin=c,width=1.1\linewidth]{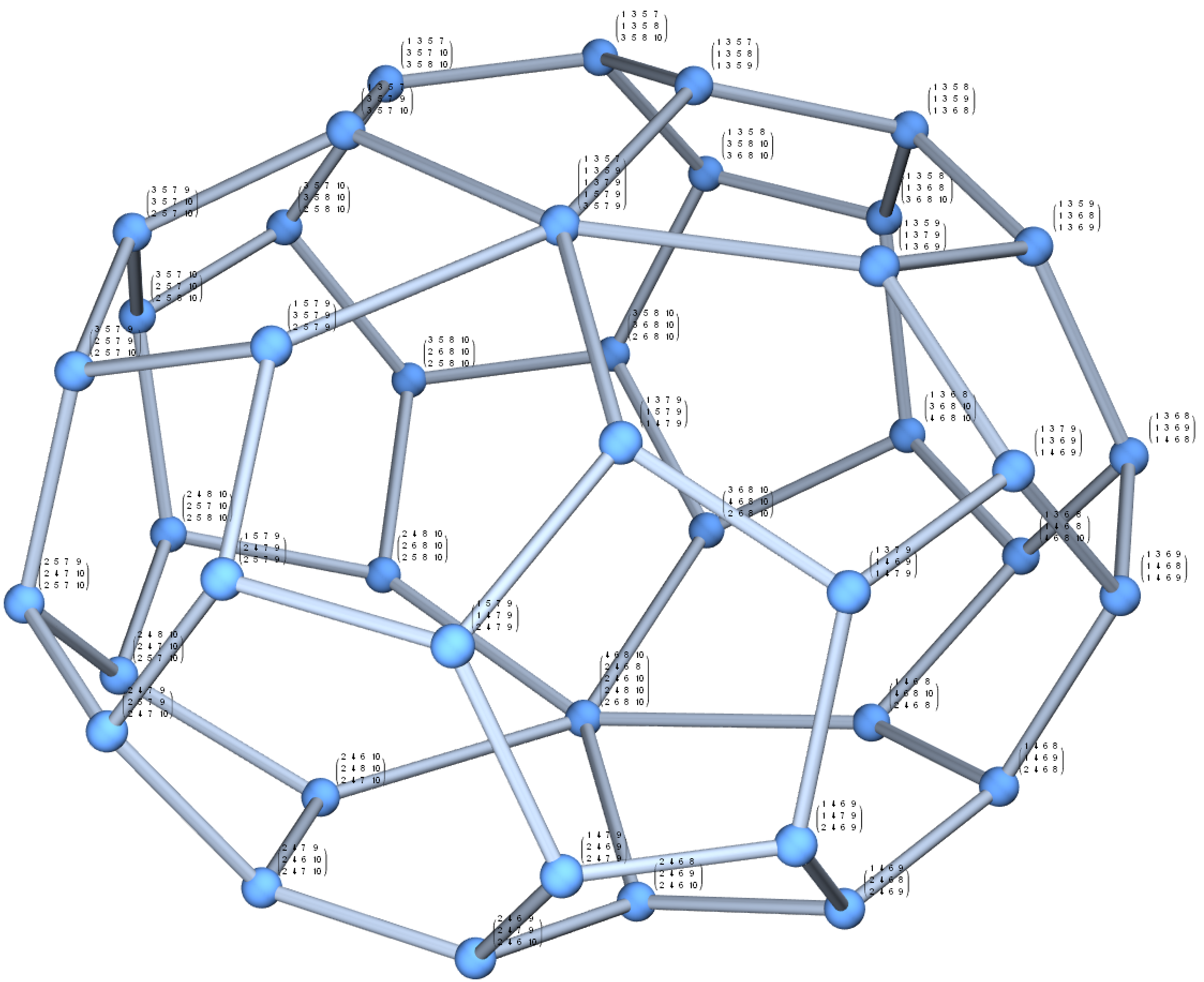}
	\caption{Condensation of the multi-split complex of $\beta_J$'s for $\Delta_{4,10}$ around 4-splits.}
	\label{fig:foursplitsg410}
\end{figure}

\end{document}